# Loss impresses human beings more than gain in the decision-making game

Jia-Quan Shen, Luo-Luo Jiang*

College of Physics and Electronic Information, Wenzhou University, Wenzhou 325035, China

*jiangluoluo@gmail.com

## Abstract:

What happen in the brain when human beings play games with computers? Here a simple zero-sum game was conducted to investigate how people make decision via their brain even they know that their opponent is a computer. There are two choices (a low or high number) for people and also two strategies for the computer (red color or green color). When the number selected by the human subject meet the red color, the person loses the score which is equal to the number. On the contrary, the person gains the number of score if the computer chooses a green color for the number selected by the human being. Both the human subject and the computer give their choice at the same time, and subjects have been told that the computer make its decision randomly on the red color or green color. During the experiments, the signal of electroencephalograph (EEG) obtained from brain of subjects was recorded. From the analysis of EEG, we find that people mind the loss more than the gain, and the phenomenon becoming obvious when the gap between loss and gain grows. In addition, the signal of EEG is clearly distinguishable before making different decisions. It is observed that significant negative waves in the entire brain region when the participant has a greater expectation for the outcome, and these negative waves are mainly concentrated in the forebrain region in the brain of human beings.

## 1、Introduction

At present, a series of research results have been obtained by using EEG to study behavioral decision-making [1-8]. There is a more profound recognition about decision-making in neural systems, especially those concerned with reward and punishment [9-11]. These studies have made considerable progress and through the system of rewards and punishments experiments to extract the features in EEG [12-16]. In addition, human behaviors can be more intensively by the change of reward and punishment mechanism, such as conditioned reinforcers and probability change [26-32]. Moreover, people often change their behavior to avoid monetary losses [33-38]. In particular, the EEG can be more powerfully influenced by people's expectations, the different in the expected value of the outcomes will also have different effects on EEG [39-43]. In ERP (event-related potential) studies on outcome evaluation or feedback processing, it has found MFN (medial-frontal negativity) is particularly sensitive to the valence of reward or performance, which is a negative deflection at frontocentral recording sites that reaches maximum between 250 and 300ms post-onset of feedback stimulus [17-21]. The MFN is more sensitive for negative feedback affected by bad outcomes, such as error responses or pecuniary losses, than for positive feedback [21-25].

Previous studies on neural activity related to losses and gains mainly concerned with after the outcomes presented. For example, studies based on scalp recording and neuro-imaging have shown that information after the outcomes presented, such as the response to gains and losses and the effects on different regions of the brain [44-47]. In fact, the EEG before the outcomes presented is also worthy of our attention. It is mainly related to people's expectations of the outcomes. It will produce different EEG due to the different expectations of the outcomes.

The main purpose of our study is to provide further evidence for the impacts of reward valence, reward magnitude and magnitude expectancy upon the EEG in outcome evaluation. Our research not only concerned about the impact of outcomes on the EEG, but also talking more attention about the decision making process impact on EEG. Here, we conducted three experiments under different reward magnitude. In every experiment, we studied the EEG which generated after the outcomes presented and before making a decision respectively.

## 2、Experimental Scheme

12 subjects participate in the experiment, including 7 boys and 5 girls. The subjects were undergraduate or graduate students with an average age of 21 years old.The subjects were selected according to the psychological experiment standard, healthy, right hand and so on, and all the subjects were voluntarily participating in these experiments. The subjects will receive a certain reward after the experiment was completed.

EEG acquisition devices use NeuroScan40 amplifiers and wearable device electrode caps. We use curry7 software to collect and process EEG signals.The stimulation of the subjects and the choice of the strategies of the subjects were carried out by using psychological experiment software E-prime.

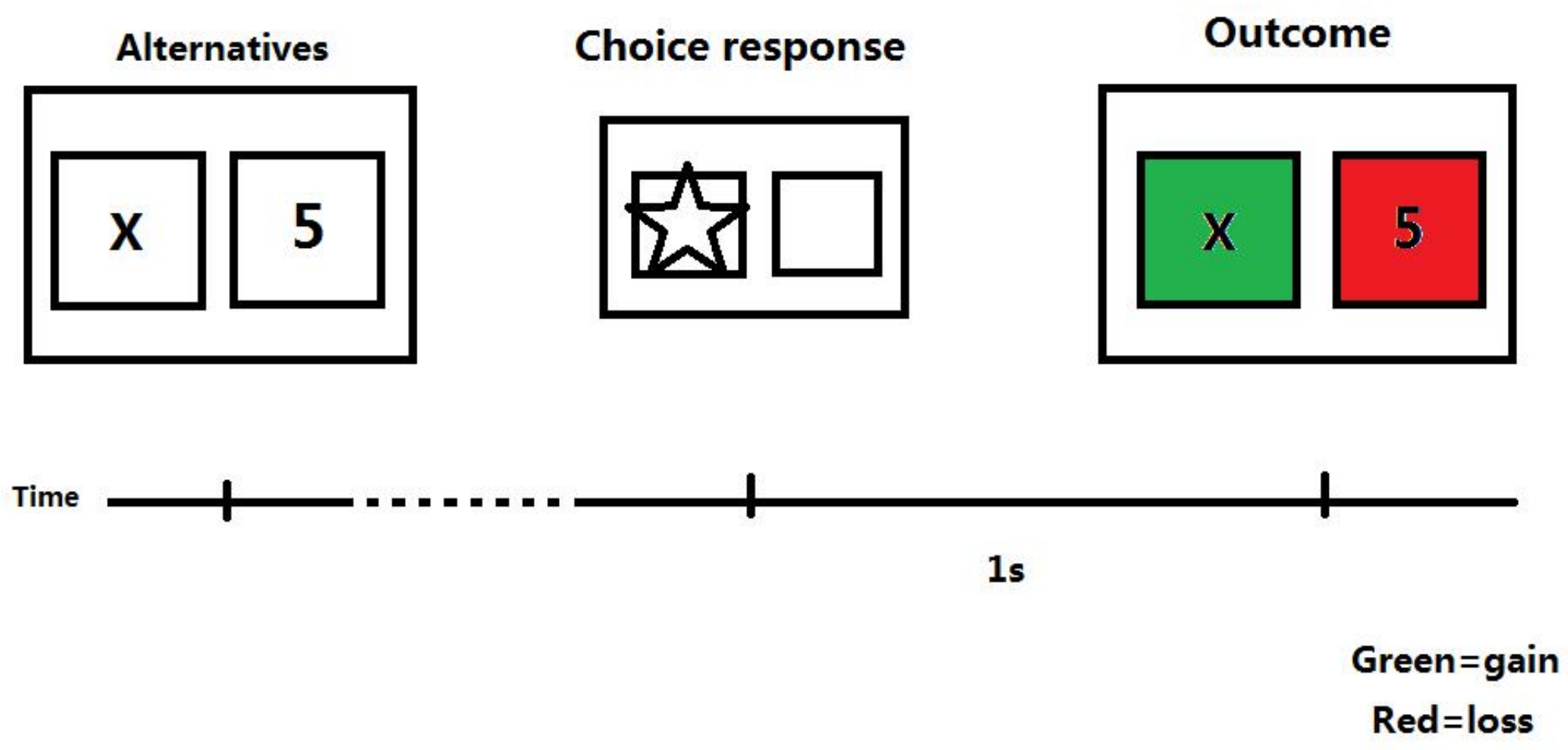

Fig.1. The experiment of game decision making. There are two squares with no color which the participants can select one of them. The squares contained the numeral 5 or X. The value of X is variable and will vary with the experiment. The squares will turn red or green when the participant makes a choice after a second. If the chosen squares turn red,it means that loss amount of money; if the squares turn green it means gain.

The subjects will be carried out three experiments, every experiment for about 15 minutes. It will be given sufficient time to rest and relax during each interval for the subjects. The schematic diagram of the experiment is shown in Figure 1. We carried out three experiments, in the first experiments x=25, corresponding in the second and third experiments the x=35 and 50 respectively. Our experiment more like a gambling task. In the experiment, participants viewed two squares with no color, each of which contained the different number, the left square contained the numeral 5 and the right square contained the numeral X. The value of X is variable and will vary with the experiment. Here we introduced a reward parameter E . It is equal to the value of left square divide by the value of right. In fact, in our experiments it means that the value of X divide by 5. The parameter E provided a systematic measure of rewards and punishments. In our first experiment the parameter E is 5, corresponding in the second and third the parameters E are 7 and 10 respectively. In particular, the parameter E also reflects the degree of risk, when the value of the parameter E in a larger range it indicates that the risk is higher.

$$E = \frac{Left}{Right} = \frac{X}{5} \quad (1)$$

Participants chose one of the squares by pressing the corresponding button. When the participants click the left mouse button it means that he choose the square on the left, clicking the right mouse button means choose the right square. One second after the choice, the color of the square will turn red or green, if the square which selected turn green it represents the income and the red means the loss. The square which participants did not choose turned red or green at the same moment that the chosen square turned red or green. For example, when the participant choose the square on the right it means that the stake is 5 yuan( RMB ), If the chosen square turns green, indicating a gain of 5 yuan, if the chosen square turns red, indicating a loss of 5 yuan. The total revenue of the participant is irrelevant to the unselected square whether it turns red or green. When the outcomes(the square turns red or green) are presented, the participant can see not only the gain or loss which he selected, but also the gain or loss of the unselected square. All the squares turn red or green with the same probability, but the participants did know it. The final score of the participants will determine how much the reward after the end of the experiment, so that the subjects can be more realistic to raise the EEG signal. The second experiment and the third experiment are similar to the first, just change the X to change 35 and 50 respectively. We collected the EEG signals of the subjects in the entire experimental.

## 3、Results

The medial frontal cortex is the main part of the brain involved in decision-making, which controls the individual's social behavior, emotions, and decision-making behavior. The Fz lead is mainly used to collect the EEG signal in medial frontal cortex. Through the analysis of the EEG signal in the Fz lead, we find that the EEG signal is different from the different outcomes, especially the gains and losses of the two cases. Moreover, the experimental results with the different of the reward parameter E are quite different.

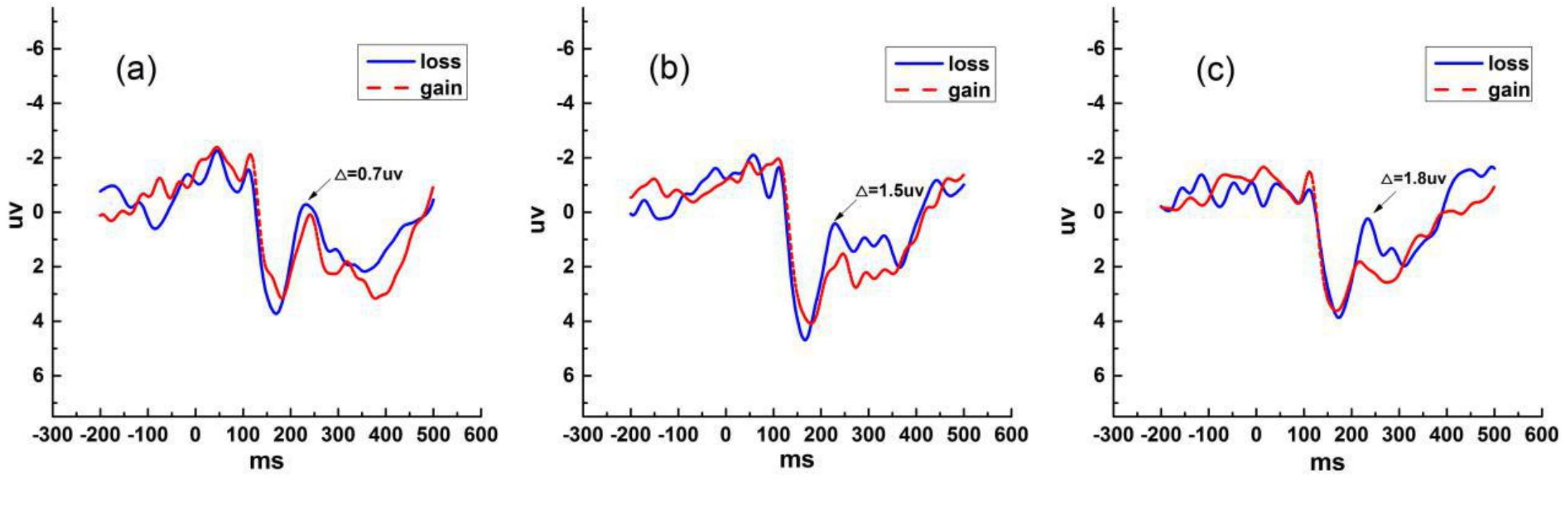

Fig.2. The EEG shown negative wave obviously in the 240ms at Fz. The solid red line corresponds to the average ERP waveform for all trials in which the participant lost money. The dashed black line corresponds to those trials in which the participant gained money. With the increase of E, the absolute value of negative wave produced by losses and gains in 240ms increased clearly. The results obtained by averaging from 12 samples. (a)The first experiments E=5, the absolute value of wave produced by losses and gains is 0.7uv. (b)The second experiments E=7, the value is 1.5uv. (c) The third experiments E=10, the value is 1.8uv.

Figure 2 shown that the EEG negative waves generated by the loss at the Fz were higher than the EEG negative waves generated by the gains at the time of around 240ms after the outcomes was presented, and the results were shown in the three experiments. There is an absolute value that will be generated by the negative wave caused by the loss and gain when the outcomes appear 240ms. This absolute value will increase with the reward parameter E.

The results suggest that the medial frontal cortex is a major part of brain involved is decision making and has a greater impact on economic losses or gains. The impact of losses is stronger than the gains on the brain in this region. The above results are from the Fz to analyze the impact of loss and gain on the brain. So what will be the impact of the brain waves generated by gain and loss on all areas of the brain when the outcomes shown after 240ms? For this, we studied the effects of loss and gain on the various regions of the brain.

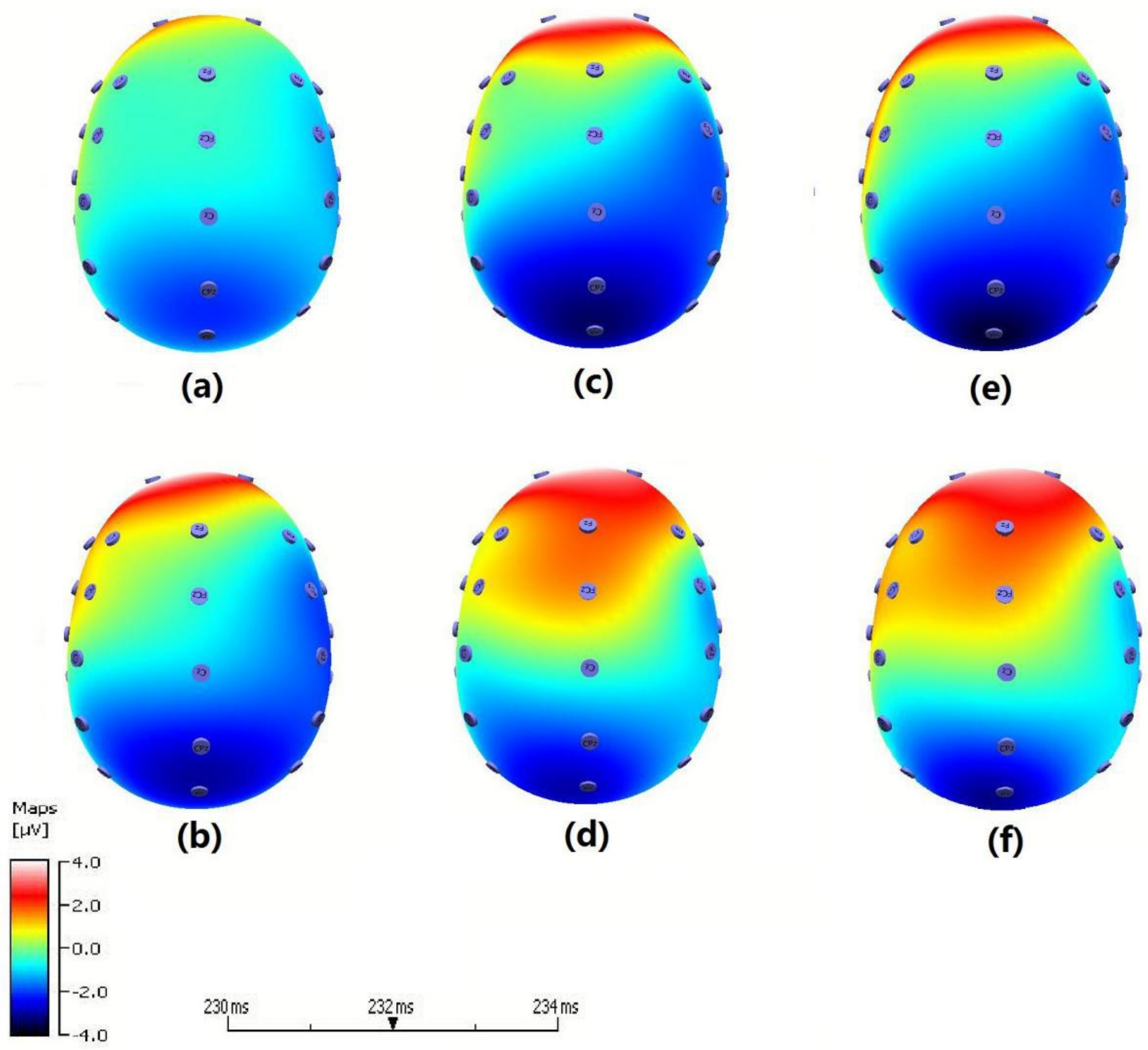

Fig.3. The effects of loss and gain on the various regions of the brain when the outcomes show after 240ms. The red region tends to show obvious positive wave characteristics, and the blue region tends to show negative wave. (a)The first experiment where E=5 and loss. (b) The first experiment where E=5 and gain. (c)The second experiment where E=7 and loss.(d)The second experiment where E=7 and gain. (e)The third experiment where E=10 and loss. (f) The third experiment where E=10 and gain.

Through the analysis of Figure 3 we found that the brain waves generated by the loss of the subject are higher than that of the gains at Fz when the outcome presented at 240ms, this phenomenon is more obvious with increase of the relative value of the loss and gain. Further observation and analysis we found that due to the impact of the loss the hindbrain area produce a more significant negative wave, and due to the impact of gain the forebrain area produce a more significant positive wave.

The medial frontal cortex of the brain is a major area of decision making is not only reflected by outcome which after the decision, but also in the decision making process and throughout the entire process of decision making. We found that the brain waves are different at Fz when before the subjects to make different choice. It is means that when the subjects are going to choose the square on the left or on the right, the brain waves are different. It is shown in Figure 4.

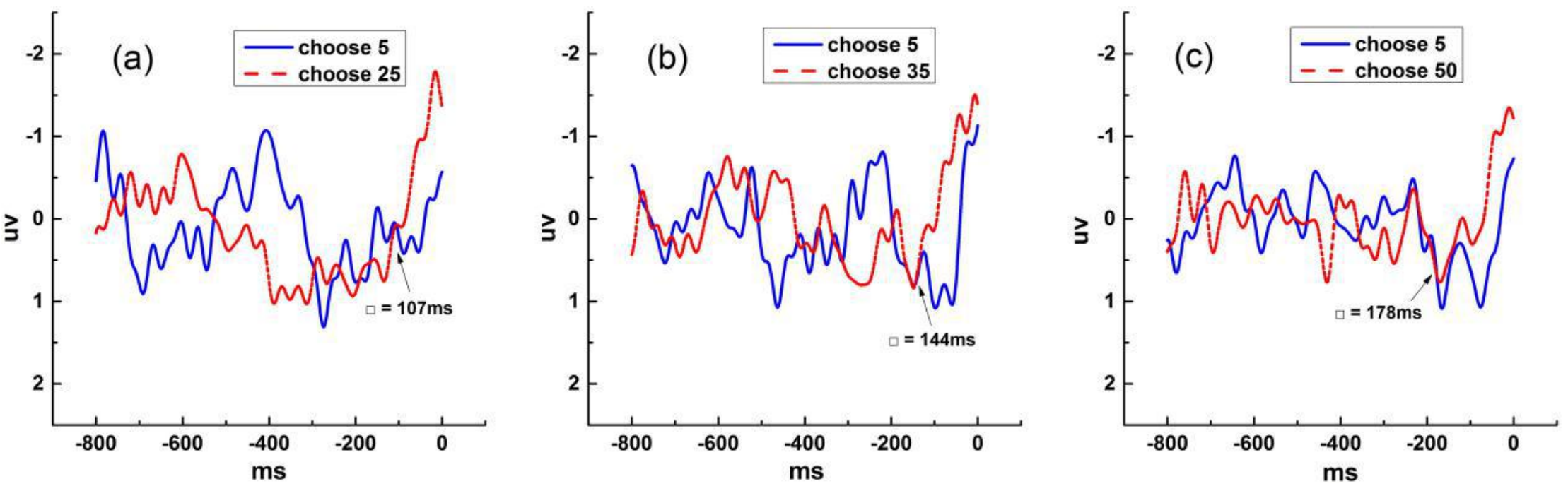

Fig.4. The EEG are completely separated from 100ms before making different decisions. The dashed red line corresponds to the average EEG waveform for all trials in which the participant will choose the square on the left (big bet x). The dashed black line corresponds to those trials in which the participant will choose the square on the right (small bet 5). (a)The first experiment where E=5. (b)The second experiment where E=7. (c) The third experiment where E=10.

As shown in Figure 4, we find that there is a big difference in the brain waves of the Fz in the 100ms before the decision is made. When the subject is going to choose the square on the left, which will choose a big bet, the negative wave of the brain wave is higher than that choose a small bet at the Fz in the 100ms before the decision is made. Moreover, the time when the brain waves are clearly separated showing a trend of gradual advance with the increase of the reward parameter E. In the first experiment where the E=5, the time of the last intersection of the EEG is in 107ms before making the decision. While in the second experiment where the E=7 the time is 144ms, and in the third experiment where the E=10 the time is 178ms.

When people making decisions they always want the choice which they choose will get the gains. When the participant chooses the square on the left it means that they make a big bet, at this time the participant has a larger expectation for the outcome. In contrast, when the participant chooses the square on the right it means that they make a small bet, at this time the participant has a smaller expectation for the outcome. Because of the different expectations of the outcomes, the brain waves in the medial frontal cortex are also different, when they have greater expectations of what they choose, they can produce a larger negative wave that the smaller ones in this brain area. In our three experiments, the participants have the same expectations when they choose the square on the right, but when they choose the square on the left they will have different expectations, the participants will have the biggest expectations when they choose the square on the left in the third experiment which the E=10. We find that the negative waves produce at the Fz can be separated from the different expectations earlier if the participants have greater expectation.

From Figure 4 ,we can see that the absolute value of the brain waves produced by different expectations before the decision 50ms reaches a maximum value, that is to say, the different of brain waves of brain waves produced bu different expectations is most obvious in this time. Therefore, we investigate the effects of different expectations on all regions of the brain in this time, to further explore the impact of behavioral decisions on brain waves.

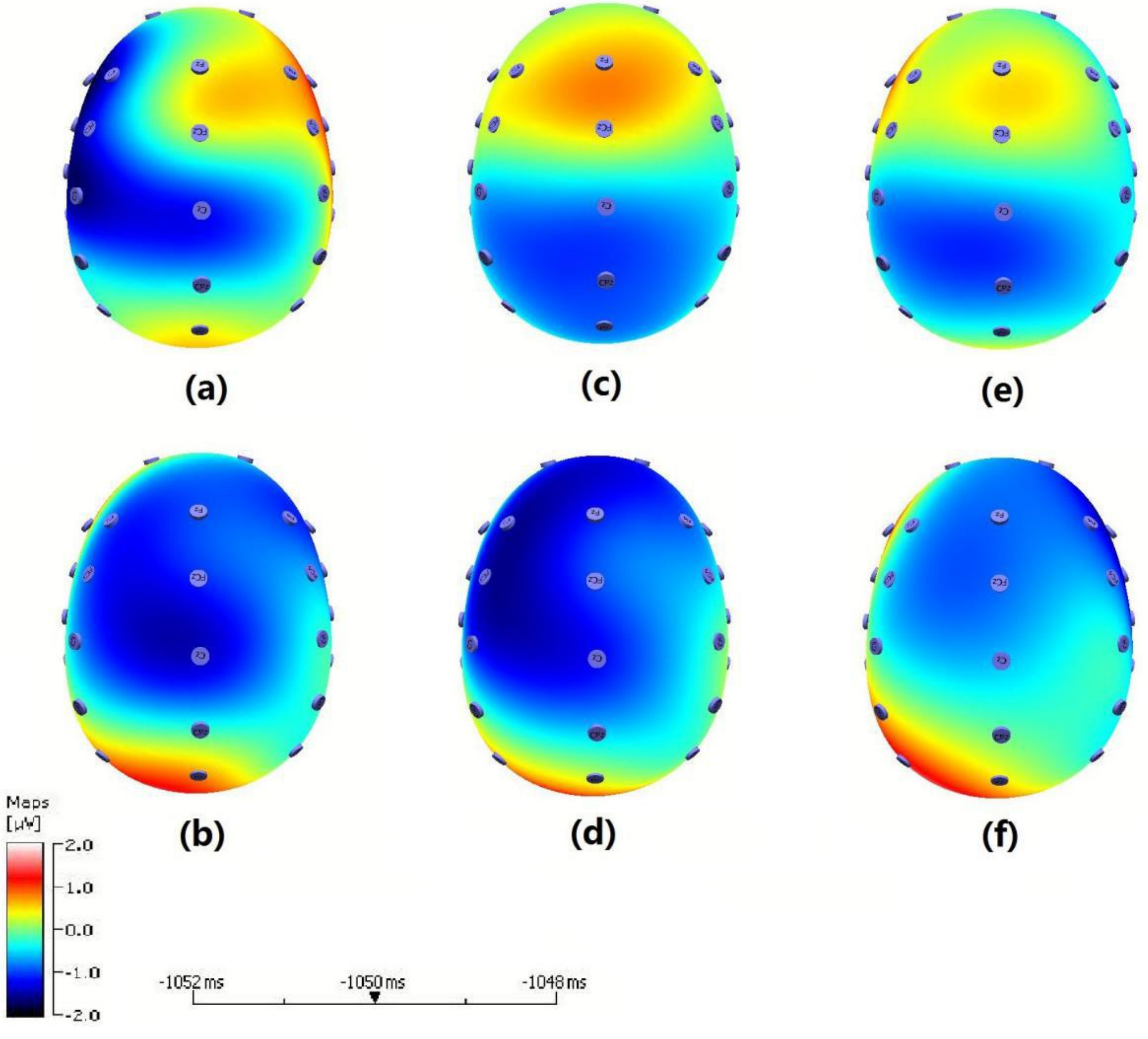

Fig.5.The effects of different expectations on the various regions of the brain when 50ms before making a choice. The red region tends to show obvious positive wave characteristics, and the blue region tends to show negative wave. (a)The first experiment where E=5 and choose 5. (b)The first experiment where E=5 and choose 25. (c)The second experiment where E=7 and choose 5.(d)The second experiment where E=7 and choose 35. (e)The third experiment where E=10 and choose 5.(f)The third experiment where E=10 and choose 50.

Figure 5 shows that the entire brain area produces a negative wave is more obvious when the subjects have greater expectations in the same experiment, and these negative waves are mainly concentrated in the forehead frontal region. When people choose different bets will have different expectations for the outcome, so the choice that the participant will make before making the selection will have a different effect on the brain waves.

## 4、Conclusion

Overall, our results show that the medial frontal cortex is the main part of the brain involved in decision-making, the impact of losses on the brain is more intense in this area. It is found that due to the impact of the loss the hindbrain area produce a more significant negative wave, and due to the impact of gain the forebrain area produce a more significant positive wave. Because of the influence of different expectations, the brain waves are also different in the forebrain region, which produce large negative waves in this area when the participant have a greater expectation for the outcome which they choose. Furthermore, it will generate significant negative waves in the entire brain region when the participants have a greater expectation for the outcome, and these negative waves are mainly concentrated in the forebrain region. In addition, signals of EEG are clearly distinguishable before making different decisions. So we can make a prediction for the upcoming select on the basis of these distinguished features in our next work.